\documentstyle[amssymb,12pt]{article}

\input{tcilatex}
\begin{document}

\begin{titlepage}
\begin{center}
{\hbox to\hsize{
\hfill \bf HIP-2001-05/TH}}
{\hbox to\hsize{\hfill  }}

\bigskip
\vspace{3\baselineskip}

{\Large \bf

Proton stability in TeV-scale GUTs\\}

\bigskip

\bigskip

{\bf Archil B. Kobakhidze$^{\mathrm{a,b}}$ \\}
\smallskip

{ \small \it
$^{\mathrm{a}}$HEP Division, Department of Physics, University of Helsinki and \\
Helsinki Institute of Physics, FIN-00014 Helsinki, Finland\\
$^{\mathrm{b}}$Andronikashvili Institute of Physics,GE-380077 Tbilisi, Georgia\\}

\bigskip

\vspace*{.5cm}

{\bf Abstract}\\
\end{center}
\noindent
We discuss the proton decay problem in theories with low  gravity and/or GUT scales.
We pointed out that the gravity induced proton decay can be
indeed suppressed up to a desired level, while the GUT origin of the proton
instability is rather problematic. To solve this problem we suggest the GUT model
where the proton is stable in all orders of perturbation theory.
This can be simply achieved by the replication of quark-lepton families with ordinary
quarks and leptons residing in different GUT representations and
by an appropriate dimensional reduction. The model
predicts extra mirror states which along with the GUT particles and the excitations
of extra dimensions could be observable at high-energy colliders providing
the unification scale is in the TeV range.

\bigskip

\bigskip

\end{titlepage}

Over the few past years a great interest attract the Brane World models
where the higher-energy scales such as a Planck scale \cite{1,2} and/or GUT
scale \cite{3,sp1} are lowered down to the energies accessible for colliders%
\footnote{%
For an earlier proposal of TeV-scale extra dimensions see \cite{4}. The
higher-dimensional scales are effectively lowered in the Randall-Sundrum
Brane World as well for the observer living on the so-called TeV-brane \cite
{5}.}. Within these models new approaches to the long-standing problems of
4-dimensional particle physics such as hierarchy problem \cite{1,2,5,6},
fermion masses \cite{3,7,8,sp2} etc. have been proposed. However, as it
usually happens, these new scenarios brought new problems as well. Among
them one is the potentially large and thus phenomenologically unacceptable
violation of certain global symmetries in the low-energy  effective theory
\cite{9}. This might be problematic for the realistic models since the
strong violation of global symmetries could induce, say, large flavour
changing neutral currents, large neutrino masses, unacceptably fast proton
decay etc.  In this note we will concentrate on the proton stability problem
in GUTs with low scale unification \cite{3,sp1,10}.

Proton decay will take place if baryon ($B$) and lepton ($L$)
numbers are simultaneously violated. The typical, lowest order
(dimension 6) $B$ and $L$ violating operator can be written as:
\begin{equation}
\frac{c}{\Lambda ^{2}}\left( qqql\right) ,  \label{1}
\end{equation}
where $q$'s ere a quark fields and $l$ denote the leptons (an appropriate
contraction of the internal and Lorentz indices in (\ref{1}) is understood).
The mass scale $\Lambda $ in the denominator of (\ref{1}) is a typical
energy scale where the $B$ and $L$ violation happens and the constant $c$
accounts a particular dynamics which leads to (\ref{1}) at low-energies.

When two of $q$'s in (\ref{1}) denote up-quarks, while the third $q$ is a
down one, and $l$ is an electron, the operator (\ref{1}) can be responsible
for the proton decay into the pion and positron: $p\rightarrow \pi ^{0}e^{+}$%
. In many GUT models, for example, this is a dominant channel for the proton
decay and the current experimental limit\footnote{%
The partial life-times of proton for other channels are also in the range
given in (\ref{2}).}
\begin{equation}
\tau _{p\rightarrow \pi ^{0}e^{+}}\gtrsim 1.6\cdot 10^{33}yr  \label{2}
\end{equation}
puts strong constraints on the possible GUT extensions. Many simple
non-supersymmetric GUTs are excluded (see, however, \cite{11}) and the
simplest $SU(5)$-type supersymmetric GUTs also fail to satisfy the bound (%
\ref{2}) \cite{12}, unless some extra mechanism ensuring desired suppression
of $B$ and $L$ violating operators\footnote{%
In supersymmetric GUTs the dominant contribution to the proton decay comes
from the dimension 5 operators induced by exchange of the colored Higgsino.}
is introduced \cite{13,14}. In GUTs the mass scale $\Lambda $ in (\ref{1})
is associated with the scale of GUT-symmetry breaking, more precisely, with
the masses of GUT particles mediating $B$ and $L$ violating interactions ($%
X,Y$ gauge bosons and colored Higgs, and their superpartners in the case of
supersymmetric GUT). These masses are expected to be of the order of
unification scale, which for the supersymmetric GUTs turns out to be around $%
10^{16}$ GeV. Clearly, all conventional GUTs with $\Lambda <10^{15\div 16}$
GeV will inevitably face with the proton stability problem. Say, if the
unification scale is around TeV \cite{3,sp1,10}, the proton life-time
becomes 48$\div $52 orders of magnitude smaller than the experimental upper
bound quoted above (\ref{2}).

Moreover, there could be also another, non-GUT origin of the effective
operators (\ref{1}) responsible for the proton decay which we would like to
discuss now. It is widely discussed that the gravity itself can be
responsible for the violation of global numbers \cite{15,16,17,18,19,20,21}
including those of $B$ and $L$ \cite{16,17}. One can distinguish two
potential gravitational sources of proton decay: one is mediated by a
virtual black hole and another is that due the wormhole background. For the
conventional Planck  scale $M_{Pl}\simeq 10^{19}$ GeV the life-time of
proton decaying gravitationally is about 12 orders of magnitude less then
upper experimental limit (\ref{2}) and thus proton can be treated as
practically stable. But now, if the fundamental Plank scale is actually TeV,
rather then $10^{19}$ GeV, as it is proposed in \cite{1,2,3}, the disaster
at first glance will be inevitable \cite{22}. We would like to stress here,
however, that the conclusions of Ref. \cite{22} is by far less obvious.
Contrary to \cite{22}, with our present knowledge of black hole and wormhole
physics, one can equally well motivate, that the violation of global quantum
numbers by gravity is extremely suppressed, even if the fundamental Planck
scale is as low as a few TeV.

The typical argumentation for the proton decay by a virtual black
holes is as follows. The proton, viewing as a hollow sphere of a
radius $r\sim m_{proton}^{-1}\approx 10^{-13}$cm with three
valence quarks inside, can transmute into another species of
particles. This might happen  when, say, two of the valence
quarks fall into the black hole which subsequently evaporates due
to the Hawking radiation \cite{23}. Since the Hawking radiation
is formed outside the black hole horizon, it is totally
insensitive to the global numbers carried by quarks inside the
horizon, and recognizes their local charges only, such as
electric charge and colour. So, among the other light particles,
black hole can emit the corresponding quark-lepton pair as well
and thus the global baryon charge get lost, while the lepton
charge appears from ''nothing''. This process can be formally
(but not equivalently) viewed as the process induced by the
exchange of a
particle with mass $M_{Pl}$, and can be described by the effective operator (%
\ref{1}), where $\Lambda =M_{Pl}$. However, the picture described above is
rather questionable since the formation and subsequent evaporation of a
black hole contradicts to the basics of quantum mechanics such as quantum
mechanical determinism. This puzzle, also known as a black hole information
loss paradox, is a subject of intensive debates during many years, but it
still far from a complete resolution \cite{18}. It is rather reasonable to
think that the solution of this puzzle within a consistent theory of quantum
gravity could automatically lead to the conservation of the global charges
or to the substantial suppression of the nucleon decay rate (see e.g., \cite
{17}). So we can at least say, that so far there is no firm theoretical
ground to indisputably believe that the global symmetries are indeed
violated by a virtual black holes. Thus, the constraints such as discussed
in \cite{22} can not be viewed as a critical theoretical argument against
the models with low scale gravity \cite{1,2}.

The situation with wormholes is more clear. It was argued that the wormholes
also lead to the violation of global charge conservation \cite{19,20,21},
but unlike the black holes this does not cause the information loss paradox.
The difference is that in the case of wormholes there appear an effective
coupling constant for the transition rate from one initial pure state to a
different final pure state. These transitions look very similar to those
which take place in the instanton background. So, the operators (\ref{1})
responsible for the proton decay seems to indeed appear in the wormhole
background and the constant $c$ in (\ref{1}) is proportional to the
(non-perturbative) exponential factor
\begin{equation}
c\sim e^{-S},  \label{3}
\end{equation}
where $S$ is a wormhole action. Now, if $S\gtrsim 59$ we are perfectly safe
to satisfy the bound (\ref{2}) even if the fundamental Planck scale is
around TeV, $\Lambda \sim M_{Pl}\sim $TeV. So, the question is: can we
naturally obtain such a large values for the wormhole action $S$? The
calculations performed in \cite{21} show that most likely this is not the
case within the pure 4-dimensional Einstein-Hilbert gravity ($S\sim {\cal O}%
(10)$). However, the wormhole action turns out to be very sensitive to the
modifications of the Einstein-Hilbert gravity. Particularly, in the case of
extra dimensions with relatively large compactification radii $%
R_{c}M_{Pl}\sim {\cal O}(10)$ the action can be as large as $S\sim {\cal O}%
(100)$\footnote{%
The main contribution to the wormhole action is proportional to $\sim
r^{2}M_{Pl}^{2}$, where $r$ is the size of the wormhole throat. In the
presence of extra dimension of radius $R_{c}$, $r$ could be of the order of $%
R_{c}$.} \cite{21}. Thus the large extra dimensions lead to the substantial
increase of the wormhole action and, as a consequence, to the dramatic
suppression of the global charge violating processes. But this is precisely
the picture we have within the  TeV scale gravity models, where at least two
of extra dimensions should have even hierarchically larger radii ($R_{c}\sim
$mm) compared to the fundamental TeV-scale in order to account the apparent
weakness of gravity in the visible 4-dimensional world \cite{1,2}. Thus one
can expect that even in the case when the fundamental Planck scale is as low
as TeV the wormhole induced non-conservation of global numbers are safely
suppressed due to the large wormhole action. From the above discussion we
conclude that gravity induced proton decay can be indeed suppressed up to a
desired level even for the fundamental Planck scale in the TeV range, while
the GUT origin of the proton instability is rather problematic.

Meanwhile, some mechanisms to suppress potentially large violation of global
charges in theories with low Planck/GUT scale have been recently suggested
\cite{2,3,8,9,24}. Basically they can be divided into two classes. One is
the ''conventional'' approach relies on the gauging of corresponding global
symmetries say that of baryon number \cite{2}, flavour symmetry \cite{9}, or
on the extra gauge symmetries invoked to ensure proton stability \cite{24}.
However, most of the proposed gauge symmetries can not be straightforwardly
extended to the GUT framework, since they treat quarks and leptons
differently. Another approach is intrinsically higher dimensional and relies
either on imposition of non-trivial boundary conditions on GUT fields within
the framework of orbifold compactification \cite{3}\footnote{%
Further recent discussions on non-trivial boundary conditions within the
orbifold compactification in connection with GUT model building and
supersymmetry breaking can be found in \cite{25} and \cite{26}, respectively.%
}, or on the localization of quark and lepton fields on a thick 3-brane at a
different points along the extra dimension \cite{8}. These scenarios, being
compatible with GUTs, require the chiral matter (i.e. ordinary quarks and
leptons) to be stuck at a 4-dimensional hypersurface (3-brane) if the
unification is assumed to be around the TeV scale. Indeed, in the scenario
of Ref. \cite{3}, for example, if one allows the third family of quarks and
leptons to freely propagate in extra dimensions, then even if the light
families are localized on a 3-brane, the proton decay mediated by $X$,$Y$
bosons and colored Higgs (Higgsino) takes place due to the mixing of light
quarks with those of top and bottom. The proton stability in this case
demands the unification scale greater than $10^{12}$ GeV \cite{3}\footnote{%
In fact, even the ''minimal model'' discussed in \cite{3} is not fully
satisfactory, because while the $X$ and $Y$ gauge bosons are decoupled from
the ordinary quarks and leptons owing to the orbifold projection, their $N=2$
scalar partners do propagate on a 3-brane. Being coupled with quarks and
leptons there, they also can mediate unacceptably rapid proton decay.}.

Here we propose an alternative possibility to ensure proton stability in low
scale GUTs. Namely, we will discuss below an explicit and rather simple
model based on $SU(5)$ gauge symmetry where the proton is stable in all
orders of perturbation theory\footnote{%
For earlier GUT models with absolutely stable proton see e.g.
\cite{pr}.}. Note once again that from the above discussion
follows that it is sufficient to keep global $B$ or $L$
symmetries exact in order to have a long-lived proton.

Let us consider supersymmetric $SU(5)$ model in higher-dimensional
space-time. For simplicity we restrict our discussion by the case of one
compact extra space-like dimension with a certain topology. Since we are
considering a supersymmetric theory in five dimensions, we have the model
invariant under the $N=2$ supersymmetry in the flat uncompactified limit,
and thus it is vector-like. Now, depending on the topology of extra
dimension and transformation properties of the superfields involved, the
theory can be arranged in such a way that after dimensional reduction one
gets either $N=2$, or $N=1$ supersymmetric theory or even a
non-supersymmetric theory. In the later cases one can also project mirror
states thus obtaining a chiral theory in four dimensions.

As usually, each family of ordinary quarks and leptons is placed in $%
\overline{5}$ and $10$ irreducible representations of the $SU(5)$ group but
now the ordinary quarks and leptons are supplemented by the mirror states.
The mirror states combine with ordinary quarks and leptons to  form $N=2$
hypermultiplets:
\begin{equation}
{\cal Q}=\left( Q_{L},Q_{R}\right) \sim \overline{5},  \label{4}
\end{equation}
\begin{equation}
{\cal D}=\left( D_{L},D_{R}\right) \sim 10,  \label{5}
\end{equation}
Here $Q_{L}(D_{L})$ and $Q_{R}(D_{R})$ are $N=1$ left-handed and
right-handed chiral quintuplet (decuplet) superfields, respectively, the
scalar components (as well as auxiliary fields) of which form doublets of a
global $SU(2)_{R}$ $R$-symmetry of $N=2$ supersymmetry. The decomposition of
the fields in (\ref{4},\ref{5}) under the $SU(3)_{C}\otimes SU(2)_{W}\otimes
U(1)_{Y}$ Standard Model gauge group is familiar:
\begin{equation}
Q_{L(R)}=\overline{d}_{L(R)}\sim \left( \overline{3},1,-\sqrt{\frac{2}{15}}%
\right) +l_{L(R)}\sim \left( 1,\overline{2},\sqrt{\frac{3}{20}}\right)
\label{6}
\end{equation}
\begin{eqnarray}
D_{L(R)} &=&\overline{u}_{L(R)}\sim \left( \overline{3},1,\frac{2}{\sqrt{15}}%
\right) +q_{L(R)}\sim \left( 3,2,-\frac{1}{\sqrt{60}}\right) +  \nonumber \\
\overline{e}_{L(R)} &\sim &\left( 1,1,-\frac{3}{\sqrt{15}}\right) .
\label{7}
\end{eqnarray}
Note that so far there is no symmetry at hand which suppresses the direct
mass terms for the quintuplets (\ref{4}) $M_{{\cal Q}}$ and decuplets (\ref
{5}) $M_{{\cal D}}$ and a priory these masses can be as large as the
fundamental Planck scale $M_{Pl}$.

$N=2$ $SU(5)$ vector supermultiplet
\begin{equation}
{\cal V}=\left( V,\Phi \right)  \label{8}
\end{equation}
contains $N=1$ 4-dimensional gauge superfield $V=\left( A^{\mu },\lambda
^{1},X^{3}\right) $ as well as $N=1$ chiral superfield $\Phi =\left( \Sigma
+iA^{5},\lambda ^{2},X^{1}+iX^{2}\right) $ both in the adjoint
representation of the $SU(5)$ group ($V=V^{a}T^{a}$, $\Phi =\Phi ^{a}T^{a}$,
where $a=1,...,24$ and $T^{a}$ are the $SU(5)$ generators in fundamental
representation, $TrT^{a}T^{b}=\frac{1}{2}\delta ^{ab}$) with the following
decomposition:
\begin{eqnarray}
V &=&G\sim \left( 8,1,0\right) +W\sim \left( 1,3,0\right) +X\sim \left(
\overline{3},2,-\sqrt{\frac{5}{12}}\right) +  \nonumber \\
\overline{X} &\sim &\left( 3,2,\sqrt{\frac{5}{12}}\right) +S\sim (1,1,0)
\label{9}
\end{eqnarray}
\begin{eqnarray}
\Phi &=&\Phi _{G}\sim \left( 8,1,0\right) +\Phi _{W}\sim \left( 1,3,0\right)
+\Phi _{X}\sim \left( \overline{3},2,-\sqrt{\frac{5}{12}}\right) +  \nonumber
\\
\Phi _{\overline{X}} &\sim &\left( 3,2,\sqrt{\frac{5}{12}}\right) +\Phi
_{S}\sim (1,1,0)  \label{10}
\end{eqnarray}
The gaugino $\lambda ^{1}$ and the Higgsino $\lambda ^{2}$ form an $%
SU(2)_{R} $ doublet, while auxiliary fields $X^{1,2,3}$ transform as $%
SU(2)_{R}$ triplet. Finally, we also need to introduce at least one $SU(5)$
fundamental and one anti-fundamental hypermultiplets:
\[
{\cal H}=(H_{L},H_{R})\sim 5,
\]
\begin{equation}
H_{L(R)}=h_{L(R)}^{C}\sim \left( 3,1,\sqrt{\frac{2}{15}}\right)
+h_{L(R)}^{W}\sim \left( 1,2,-\sqrt{\frac{3}{20}}\right) ,  \label{11}
\end{equation}
\[
\widetilde{{\cal H}}=(\widetilde{H}_{L},\widetilde{H}_{R})\sim \overline{5},
\]
\begin{equation}
\widetilde{H}_{L(R)}=\widetilde{h}_{L(R)}^{C}\sim \left( \overline{3},1,-%
\sqrt{\frac{2}{15}}\right) +\widetilde{h}_{L(R)}^{W}\sim \left( 1,2,\sqrt{%
\frac{3}{20}}\right) ,  \label{11_1}
\end{equation}
respectively, where the electroweak Higgs doublet (anti-doublet) $%
h_{L(R)}^{W}$ ($\widetilde{h}_{L(R)}^{W}$) presumably resides. ${\cal H}$
and $\widetilde{{\cal H}}$ could also have $SU(5)$-invariant masses $M_{%
{\cal H}}$ and $M_{\widetilde{{\cal H}}}$, respectively. One obvious
advantage of the $N=2$ supersymmetric GUTs is that the gauge fields in $V$
and scalars in $\Phi $ are unified in the same $N=2$ vector supermultiplet (%
\ref{8}). The scalar component of the chiral superfield $\Phi $ can be used
to break $SU(5)$ gauge symmetry down to the $SU(3)_{C}\otimes
SU(2)_{W}\otimes U(1)_{Y}$ so we do not need to introduce extra adjoint
hypermultiplet. Certainly, in this respect, the $N=2$ models are much more
predictive than the ordinary $N=1$ $SU(5)$. Alternatively, one can break $%
SU(5)$ invariance through the orbifold compactification \cite{3,25}. Having
determined the particle content of the model, an $SU(5)$ and $N=2$ invariant
5-dimensional Lagrangian can be written down as \cite{a1}\footnote{%
We consider here the model with minimal (quadratic) prepotential on the
Coulomb branch. Such a minimal prepotential can be forced by the topology of
compactified space \cite{26}.}:
\begin{equation}
{\cal L}={\cal L}^{gauge}+{\cal L}^{matter},  \label{12}
\end{equation}

\begin{equation}
{\cal L}^{gauge}=\frac{1}{4g_{GUT}^{2}}\int d^{2}\theta {\cal W}^{2}+h.c.+%
\frac{1}{g_{GUT}^{2}}\int d^{4}\theta \left( \partial _{5}V-\frac{1}{\sqrt{2}%
}(\Phi +\Phi ^{+})\right) ^{2},  \label{13}
\end{equation}
\begin{equation}
{\cal L}^{matter}=\int d^{4}\theta \left( \Psi _{L}^{+}\widehat{K}\Psi
_{L}+\Psi _{R}\widehat{K}^{-1}\Psi _{R}^{+}\right) +\int d^{2}\theta \Psi
_{R}^{+}\widehat{M}\Psi _{L}+h.c,  \label{14}
\end{equation}
where ${\cal W}$ is an ordinary strength superfield of the $SU(5)$ gauge
field, $g_{GUT}$ is an unified gauge coupling constant and
\begin{equation}
\Psi _{L(R)}=\left( Q_{L(R)},D_{L(R)},H_{L(R)},\widetilde{H}_{L(R)}\right)
^{T},  \label{15}
\end{equation}
\begin{equation}
\widehat{K}=diag\left( e^{-V_{{\cal Q}}},e^{-V_{{\cal D}}},e^{-V_{{\cal H}%
}},e^{-V_{\widetilde{{\cal H}}}}\right) ,  \label{16}
\end{equation}
\begin{equation}
\widehat{M}=diag\left( \partial _{5}+M_{{\cal Q}}-\frac{1}{\sqrt{2}}\Phi _{%
{\cal Q}},\partial _{5}+M_{{\cal D}}-\frac{1}{\sqrt{2}}\Phi _{{\cal D}%
},\partial _{5}+M_{{\cal H}}-\frac{1}{\sqrt{2}}\Phi _{{\cal H}},\partial
_{5}+M_{\widetilde{{\cal H}}}-\frac{1}{\sqrt{2}}\Phi _{\widetilde{{\cal H}}%
}\right) .  \label{17}
\end{equation}
Here $V_{{\cal Q},{\cal D},{\cal H},\widetilde{{\cal H}}}(\Phi _{{\cal Q},%
{\cal D},{\cal H},\widetilde{{\cal H}}})=V^{a}(\Phi ^{a})T_{{\cal Q},{\cal D}%
,{\cal H},\widetilde{{\cal H}}}^{a}$, and $T_{{\cal Q},{\cal D},{\cal H},%
\widetilde{{\cal H}}}^{a}$ are the $SU(5)$ generators where subscript refers
to the corresponding representation. The usual repetition of quark-lepton
generations residing in $Q_{L(R)},D_{L(R)}$ is also assumed in the above
equations.
\begin{table}[t] \centering%
\begin{tabular}{lll}
\hline
Fields &  & $Z_{2}$ parity, ${\cal P}$ \\ \hline
Vector supermultiplet, ${\cal V}$ & $\left\{
\begin{array}{l}
G(\Phi _{G})\sim \left( 8,1,0\right) \\
W(\Phi _{W})\sim \left( 1,3,0\right) \\
S(\Phi _{S})\sim \left( 1,1,0\right) \\
X,(\Phi _{X})\sim \left( \overline{3},2,-\sqrt{\frac{5}{12}}\right) \\
\overline{X}(\Phi _{\overline{X}})\sim \left( 3,2,\sqrt{\frac{5}{12}}\right)
\end{array}
\right. $ & \multicolumn{1}{c}{$
\begin{array}{c}
+(-) \\
+(-) \\
+(-) \\
-(+) \\
-(+)
\end{array}
$} \\ \hline
Quintuplets, ${\cal Q}$, $\widetilde{{\cal Q}}$ & $\left\{
\begin{array}{c}
\overline{d}_{L(R)},\widetilde{\overline{d}}_{L(R)}\sim \left( \overline{3}%
,1,-\sqrt{\frac{2}{15}}\right) \\
l_{L(R)},\widetilde{l}_{L(R)}\sim \left( 1,\overline{2},\sqrt{\frac{3}{20}}%
\right)
\end{array}
\right. $ & \multicolumn{1}{c}{$
\begin{array}{l}
-(+),+(-) \\
+(-),-(+)
\end{array}
$} \\ \hline
Decuplets, ${\cal D}$, $\widetilde{{\cal D}}$ & $\left\{
\begin{array}{c}
\overline{u}_{L(R)},\widetilde{\overline{u}}_{L(R)}\sim \left( \overline{3}%
,1,\frac{2}{\sqrt{15}}\right) \\
q_{L(R)},\widetilde{q}_{L(R)}\sim \left( 3,2,-\frac{1}{\sqrt{60}}\right) \\
\overline{e}_{L(R)},\widetilde{\overline{e}}_{L(R)}\sim \left( 1,1,-\frac{3}{%
\sqrt{15}}\right)
\end{array}
\right. $ & \multicolumn{1}{c}{$
\begin{array}{c}
-(+),+(-) \\
+(-),-(+) \\
-(+),+(-)
\end{array}
$} \\ \hline
Higgs hypermultiplets, ${\cal H},\widetilde{{\cal H}}$ & $\left\{
\begin{array}{c}
h_{L(R)}^{C}\sim \left( 3,1,\sqrt{\frac{2}{15}}\right) \\
h_{L(R)}^{W}\sim \left( 1,2,-\sqrt{\frac{3}{20}}\right) \\
\widetilde{h}_{L(R)}^{W}\sim \left( 1,2,\sqrt{\frac{3}{20}}\right) \\
\widetilde{h}_{L(R)}^{C}\sim \left( \overline{3},1,-\sqrt{\frac{2}{15}}%
\right)
\end{array}
\right. $ & \multicolumn{1}{c}{$
\begin{array}{l}
-(+) \\
+(-) \\
+(-) \\
-(+)
\end{array}
$} \\ \hline
\end{tabular}
\caption{An intrisic parity ${\cal P}$  of various fields (see eqs.
(\ref{4}-\ref{11}).}%
\end{table}%

The above Lagrangian (\ref{12},\ref{13},\ref{14}) describes the minimal
extension of the ordinary 4-dimensional $N=1$ supersymmetric $SU(5)$ GUT to
the case of five space-time dimensions. Clearly, the above model in many
aspects is unacceptable from the phenomenological point of view. Then a
certain dimensional reduction should provide the projection out of unwanted
states at low-energies in lower dimensions. We will consider here the
compactification of extra fifth dimension on an $S^{1}/Z_{2}$ orbifold. What
we will require additionally is the conservation of $B$ and/or $L$ global
charges upon the compactification. In fact this is not difficult to achieve.
First, in addition to the particle content given above, let us introduce an
extra quintuplet $\widetilde{{\cal Q}}$ $\sim \overline{5\mbox{ }}$and an
extra decuplet $\widetilde{{\cal D}}$, $\sim 10$ of matter fields per each
family of quarks and leptons. The second step is to appropriately project
the different states in ${\cal Q}$, $\widetilde{{\cal Q}}$, ${\cal D}$, and $%
\widetilde{{\cal D}}$ upon the dimensional reduction. This can be
done by assigning different $Z_{2}$ orbifold $Z_{2}$-numbers to
the quarks and leptons (and
their mirrors) in ${\cal Q}$, $\widetilde{{\cal Q}}$, ${\cal D}$, and $%
\widetilde{{\cal D}}$ as we will demonstrate just below.

An intrinsic $Z_{2}$ parity ${\cal P}$ ($=\pm 1$) of a generic 5-dimensional
bulk field $\varphi (x^{\mu },y)$ is defined as:
\begin{equation}
\varphi (x^{\mu },-y)={\cal P}\varphi (x^{\mu },y)  \label{18}
\end{equation}
Parity odd $\varphi _{+}$ (${\cal P}=1$) and parity even $\varphi _{-}$ ($%
{\cal P}=-1$) can be expanded as:
\begin{equation}
\varphi _{+}=\sum_{n=0}^{\infty }\varphi ^{(n)}(x^{\mu })\cos \left(
ny/R_{c}\right)  \label{19}
\end{equation}
\begin{equation}
\varphi _{-}=\sum_{n=1}^{\infty }\varphi ^{(n)}(x^{\mu })\sin \left(
ny/R_{c}\right) ,  \label{20}
\end{equation}
where $n$ is an integer number determining the quantized momentum
corresponding to the extra compact dimension with radius $R_{c}$. A
remarkable property of the compactification on $S^{1}/Z_{2}$ is that the
parity-odd fields $\varphi _{-}$ (\ref{20}) lack zero modes. Now let us
determine $Z_{2}$- numbers of various fields involved in our model as it is
given in Table 1. Then it can be straightforwardly checked that the
Lagrangian (\ref{12}) is actually invariant under the orbifold parity
transformations as it should be\footnote{%
Note that the orbifold symmetry forbids the mass terms $M_{{\cal Q}}$, $M_{%
{\cal D}}$ and $M_{{\cal H}}$ in (\ref{14}). Also since $\Phi _{S}$ is $%
Z_{2} $-odd it can not acquire non-zero vacuum expectation value. The GUT
symmetry is actually broken by the non-trivial boundary conditions (\ref{18}%
) determined by $Z_{2}$-numbers from Table 1.}. According to the expansion (%
\ref{19},\ref{20}) the wave functions of the parity-odd fields vanish at the
orbifold fixed-points ($y=0,\pi $) and only parity-even fields can propagate
on the 4-dimensional boundary walls. This suggest the identification of
ordinary quarks and leptons with ${\cal Q}$,$\widetilde{{\cal Q}}$, ${\cal D}
$ $\widetilde{{\cal D}}$ fragments as:
\begin{equation}
\widetilde{\overline{d}}_{L},\widetilde{\overline{u}}_{L},q_{L},l_{L},%
\widetilde{\overline{e}}_{L}.  \label{21}
\end{equation}
There also present their mirror states on the boundary wall:
\begin{equation}
\overline{d}_{R},\overline{u}_{R},\widetilde{q}_{R},\widetilde{l}_{R},%
\overline{e}_{R}.  \label{22}
\end{equation}
The gauge symmetry on the wall is just $SU(3)_{C}\otimes SU(2)_{W}\otimes
U(1)_{Y}$ one and $N=2$ supersymmetry is reduced to the $N=1$. The vector
superfields $X,\overline{X}$ from (\ref{9}) as well as chiral superfields $%
\Phi _{G},\Phi _{W},\Phi _{S}$ from (\ref{10}) are projected out since all
they are $Z_{2}$-odd. In the same way coloured triplet (anti-triplet)
superfield $h_{L}^{C}$ ($\widetilde{h}_{L}^{C}$) decouples from the
electroweak doublet (anti-doublet) $h_{L}^{W}$ ($\widetilde{h}_{L}^{W}$) on
the boundary wall, thus realizing intrinsically higher-dimensional mechanism
for the doublet-triplet splitting. Beside the mirror states (\ref{22}) we
have some additional states beyond the usual particle content of the minimal
supersymmetric Standard Model (MSSM). They are chiral leptoquark superfields
$\Phi _{X}$ and $\Phi _{\overline{X}}$ from (\ref{10}) and the colored
triplet and anti-triplet anti-chiral superfields $h_{R}^{C}$, $\widetilde{h}%
_{R}^{C}$ from (\ref{11}) and (\ref{11_1}), respectively.

Now, since the gauge superfields $X,\overline{X}$ are $Z_{2}$-odd, they
decoupled from the zero modes of quarks and leptons and thus they can not be
responsible for the $B$ and $L$ violating interactions among them anymore.
The adjoint scalar superfields $\Phi _{X}$ and $\Phi _{\overline{X}}$,
contrary, have a zero modes and couple to the matter on the boundary wall.
However, they can only transform the mirror quarks into the ordinary leptons
and the mirror leptons into the ordinary quarks (see eqs. (\ref{14},\ref{17}%
)). So among the possible final states along with the ordinary quarks and
leptons always will appear the mirror ones. This actually means that in the
limit of massless ordinary quarks and leptons and their mirror partners the
following global charges are separately conserved:
\begin{equation}
Q_{1}=B+L_{M},  \label{23_1}
\end{equation}
\begin{equation}
Q_{2}=L+B_{M},  \label{23}
\end{equation}
where the mirror baryon and lepton numbers we denote as $B_{M}$ and $L_{M}$,
respectively. To generate the masses for the ordinary quarks and leptons and
their mirror partners we add $N=2$ supersymmetry violating terms to the
Lagrangian (\ref{12}) on the boundary wall at $y=0$:
\begin{eqnarray}
{\cal L}^{Yukawa} &=&\int d^{2}\theta \delta (y)\left[ Y^{(lept.)}Q_{L}%
\widetilde{D}_{L}\widetilde{H}_{L}+Y^{(down)}\widetilde{Q}_{L}D_{L}%
\widetilde{H}_{L}+Y^{(up)}D_{L}\widetilde{D}_{L}H_{L}\right. +  \nonumber \\
&&\left. Y_{M}^{(lept.)}\widetilde{Q}_{R}D_{R}H_{L}^{+}+Y_{M}^{(down)}Q_{R}%
\widetilde{D}_{R}H_{L}^{+}+Y_{M}^{(up)}D_{R}\widetilde{D}_{R}\widetilde{H}%
_{L}^{+}+h.c.\right] ,  \label{24}
\end{eqnarray}
where $Y^{(...)}$ and $Y_{M}^{(...)}$ are the Yukawa constants for the
ordinary quarks, leptons and their mirror partners, respectively, and a
certain family structure in (\ref{24}) is assumed. Because the appearance of
$\delta (y)$-term, only the zero modes of corresponding superfields appear
in (\ref{24}). This actually means that the colored Higgs triplets from $%
H_{L}$ and $\widetilde{H}_{L}$ completely decoupled from the quarks and
leptons as well as from their mirror partners. Consequently, the
interactions in (\ref{24}) do not violate neither $B$ and $B_{M}$ nor $L$
and $L_{M}$ and $Q_{1}$ and $Q_{2}$ remain conserved\footnote{%
The mirror neutrinos have to be massive as well. For this purpose one can
introduce an extra $SU(5)$-singlet hypermultiplet ${\cal S}=(S_{L},S_{R})$
with ${\cal P}(S_{R})=-{\cal P}(S_{R})=1$. Then the coupling $\int
d^{2}\theta \delta (y)\left[ Y_{M}^{(neut)}S_{R}\widetilde{Q}%
_{R}H_{L}^{+}+h.c.\right] $ will be responsible for the
generation of the mirror neutrino mass after the electroweak
symmetry breaking. Note that the above term also preserves
$L_{M}$ and thus does not alter our discussion}. It is evident
now that, since the mirror particles is assumed to be heavier
than the ordinary ones, the proton decay is forbidden
kinematically. In other words, as long as mirror particles cannot
be produced $B$ and $L$ are separately conserved. As a result the
proton is absolutely stable. We would like to stress once again
that the above consideration is valid in all orders in
perturbation theory.

Concluding, probably one can think, that the idea of low-scale gravity and
unification is too radical to be true. What is really exciting, however, is
that this idea have indeed survived under the pressure of phenomenological
constraints. Moreover, the proposed solutions to the existing problems often
give even more exotic predictions which can be tested in the visible future.
One such model has been discussed in this paper. Motivated by the solution
of the proton decay problem we have constructed GUT model where baryon $B$
and lepton $L$ numbers are perturbatively conserved. The model predicts
extra mirror states\footnote{%
Previous models as well as phenomenology of mirror fermions can be found in
\cite{27}.} which along with the GUT particles and the excitations of extra
dimensions could be observable at high-energy colliders providing the
unification scale is in the TeV range. Thus remains to hopefully wait for
new collider experiments.

\paragraph{Acknowledgments.}

I would like to thank Zurab Berezhiani and Ilia Gogoladze for useful
discussions. This work was supported by the Academy of Finland under the
Project No. 163394.

\subparagraph{Note added}

After this work was completed the paper \cite{28} appeared in the
hep-archive where the proton decay problem is discussed within the $SU(5)$
model in five dimensions compactified on $S^{1}/(Z_{2}\times Z_{2}^{^{\prime
}})$ orbifold.

\end{document}